\documentclass[12pt]{article}
\usepackage{graphicx}
\begin{document}
\setlength{\textheight}{7.7truein} 
\thispagestyle{empty}

\centerline{\bf STORAGE CAPACITY OF EXTREMELY DILUTED HOPFIELD MODEL}
\vspace*{0.37truein}
\centerline{\footnotesize BURCU AKCAN}
\baselineskip=12pt
\vspace*{10pt}         %actual spacing
\centerline{\footnotesize Y\.{I}\u{G}\.{I}T G\"{U}ND\"{U}\c{C}}
\baselineskip=12pt
\vspace*{10pt}         %actual spacing
\centerline{\footnotesize\it Physics Engineering Department, Hacettepe 
University, Beytepe}
\baselineskip=10pt
\centerline{\footnotesize\it Ankara, 06530, Turkey}
\vspace*{10pt}         %actual spacing
\centerline{\footnotesize\it E-mail: burcu@btae.mam.gov.tr}
\centerline{\footnotesize\it E-mail: gunduc@hacettepe.edu.tr}

\vspace*{3truecm}

\centerline {\bf Abstract}
\vspace*{1 truecm}
{The storage capacity of the extremely diluted Hopfield Model is
studied by using Monte Carlo techniques. In this work, instead of
diluting the synapses according to a given distribution, the dilution
of the synapses is obtained systematically by retaining only the
synapses with dominant contributions. It is observed that by using the
prescribed dilution method the critical storage capacity of the system
increases with decreasing number of synapses per neuron reaching
almost the value obtained from mean-field calculations. It is also
shown that the increase of the storage capacity of the diluted system
depends on the storage capacity of the fully connected Hopfield Model
and the fraction of the diluted synapses.}{}{}

\vspace*{50pt}

{\bf \small keywords : Hopfield Model;  extremely diluted neural networks; 
storage capacity; Monte Carlo simulation}

\pagebreak 

\section{Introduction}         %) A SECTION HEADING
\vspace*{-0.5pt}

Statistical mechanics provides very powerful theoretical framework to
study the storage capacities and the generalization abilities for a
variety of artificial neural network architectures. One of the most
popular and well-studied artificial neural network model is the
Hopfield Model.  In the Hopfield Model\cite{1}, neurons have two
levels of activity and in the original form of model, the neurons are
connected with all other neurons with synaptic couplings. The synaptic
couplings, $J_{ij}$, between $i^{th}$ and $j^{th}$ neurons are set by
the Hebb Learning Process\cite{2} which ensures the dynamic stability
of the initially introduced patterns. The $\mu^{th}$ pattern in the
original set of patterns, ${\xi_{i}^{\mu}}$ are assumed to be random
with equal probabilities for ${\xi_{i}^{\mu}=\pm1}$.

$\;$

One can define the critical storage capacity
($\alpha_{c}$) as,

\begin{equation}
\alpha_{c}=\frac{P_{max}}{N}
\end{equation}

where $N$ is the number of neurons and $P_{max}$ is the maximum number
of patterns  that can be stored into the neuron system before the
total loss of memory\cite{3}. The critical storage capacity
($\alpha_{c}$) for the two-state fully connected neuron system is
calculated by means of mean-field and numerical simulation
methods. Typical values are
$0.1379$\cite{3} and $0.144$\cite{4} by using replica symmetry
breaking method and $0.144\pm0.009$\cite{3},
$0.145\pm0.001$\cite{5} and $0.143\pm0.001$\cite{6} by using
numerical simulation techniques.

$\;$

For the fully connected networks the critical storage capacity sets
severe restrictions on the applicability of the model to more
realistic problems of the physical systems. In this case one turns to
the biological systems where the storage capacity is extremely higher
than the above given values. In the biological neural networks, the
neurons are very sparsely connected and the connections are set by the
personal experiences. This extremely successful example led scientists
to devote a special interest to the diluted networks where only a
fraction of neurons is connected. Theoretical studies showed that the
storage capacity per neuron in diluted artificial neuron systems can
be substantially larger than the fully connected networks\cite{7,8}.
This theoretical result can intuitively be understood by considering
the local field acting on a single neuron. If the stored patterns are
very limited in number, the effective local field sets the sign of the
neuron (sign of the local field) at a given site without any
ambiguity. Increasing the number of stored patterns causes frustration
on the neurons. Hence dilution may decrease the number of frustrated
neurons, therefore increase in the storage capacity may be observed.

In this work we aim to study the storage capacity of the extremely
diluted Hopfield Model by using Monte Carlo simulation techniques. We
aim to introduce a dilution prescription which resembles the
biological learning process. In the selection of the remaining
connections the most prominent contributions to the connection matrix
are considered for each neuron. This prescription leads to an
extremely diluted system and yet the connections are still sufficient
to recover large number of stored patterns.

$\;$

The paper is organized as follows: In the next section we have introduced  
our dilution method and the thermodynamic observables in order to study memory 
effects. In the third section we have presented our simulation results and 
how the storage capacity of the diluted Hopfield network can change by using 
our dilution techniques. The final section is devoted to the discussion of 
our results and conclusion.

%%%%%%%%%%%%%%%%%%%%%%%%%%%%%%%%%%%%%%%%%%%%%%%
\section{Model and the Method}
%%%%%%%%%%%%%%%%%%%%%%%%%%%%%%%%%%%%%%%%%%%%%%

The dynamics of the two-state neuron system is defined by the Hopfield 
Hamiltonian;

\begin{equation}
\label{Hamiltonian}
H = \frac{1}{2}\sum\limits_{ij}J_{ij}S_{i}S_{j}
\end{equation}

where $S_{i}$ is the $i^{th}$ binary neuron that can take the values
$\pm1$.  We have used zero temperature dynamics to update the neurons
at a given time $t$ according to the following rule;

\begin{equation}
\label{Dynamics}
S_{i}(t+1) = Sign(\sum\limits_{i\neq j}J_{ij}S_{j}(t))
\end{equation}

and the synaptic matrix element between $i^{th}$ and $j^{th}$
neurons, $J_{ij}$, is defined as;

\begin{equation}
\label{Couplings}
J_{ij} = C_{ij}\sum\limits_{\mu =1}^{P}\xi_{i}^{\mu}\xi_{j}^{\mu}
\end{equation}

where $P$ indicates the number of initial patterns to be stored in the
system, $i$ and $j$ run from $1$ to $N$ and  $N$ is the number of
neurons in the system. The elements of the dilution matrix, $C$, define the
dilution of the synapses. The $C_{ij}$'s can take the value $''1''$ or
$''0''$ depending on whether the connection between $i^{th}$ and
$j^{th}$ neuron exits or not.

$\;$

In our study, the dilution of the couplings is realized in a guided
way. Instead of diluting the connections according to a given
distribution, we have introduced a prescription by which the dilution
of the synapses is realized. Our prescription resembles very much the
actual learning process: In the learning process some events or
patterns are easily recalled if many different events or patterns in
the past experiences have similarities or relations.  In fact in the
fully connected Hopfield Model, since the original patterns have
random orientations at the $j^{th}$ position, the coupling matrix $J$
may contain many zeros or very small random values whose contribution
to the dynamics of the $i^{th}$ neuron is only to randomize the system
at the following time steps.  The most prominent contributions to the
dynamics of the $i^{th}$ neuron are coming from the sites, $j$, at
which many different original patterns posses the same value. The
number of sites with this property rapidly decreases as the number of
initially thought patterns increases. Nevertheless, the elimination of
the synapses with random contributions will emphasize the effect of
these sites. These contributions are decisive contributions for the
orientation of the $i^{th}$ neuron at the time $t+1$ when the system
is extremely diluted. In the case of extremely diluted Hopfield Model,
many couplings will be taken as zero leaving only those that
contribute the most. Since the majority of the neighboring neurons
indicates a net orientation and hence in the summation in
Eq. (\ref{Couplings}) the cancellations are minimal.

$\;$

To implement this prescription, the strengths of the
connections between $i^{th}$ neuron and all other neurons are
calculated prior to the dilution process. In order to determine the
undiluted synapses, we have considered how many neurons at a given
site $j$ in different patterns carry the same sign with the $i^{th}$
site, $T^{same}_{j}$, and how many neurons posses opposite sign,
$T^{opposite}_{j}$. The number of like neurons in all patterns at
the location $j$

\begin{equation}
T^{same}_{j}=\sum_{\mu=1}^{P} \delta_{\xi^{\mu}_i,\xi^{\mu}_j}
\end{equation}

and the number of oppositely signed neurons at the location $j$,

\begin{equation} 
T^{opposite}_{j}=\sum_{\mu=1}^{P}(1-\delta_{\xi^{\mu}_i,\xi^{\mu}_j})
\end{equation}

constitute two sets of values. These sets are sorted separately by
value. Apart from the zero values which may exist in both sets the $j$
values can not be the same. By merging these two sets together and by
considering the fraction of the diluted neurons. Only the required
number of terms starting from the highest values are taken into
consideration. From this set one can obtain the locations ($j$'s) of
the neighboring neurons with non-vanishing synapses. The relevant
elements of the dilution matrix $C$ are set to unity starting from the
largest value from both sets i.e. only the prominent contributions are
taken into account. Even if the system is not extremely diluted or
fully connected, the same prescription still works.

$\;$

In order to  study  the time evolution of a neuron system, the 
following observables are calculated:

i) The average overlap $m(t)$ between the states $S(t)$ and the $\mu^{th}$ original pattern 
$\xi^{\mu}$ that are all stored in the memory;

\begin{equation}
m(t) = \frac{1}{N P}\sum\limits_{\mu=1}^{P}\sum\limits_{i=1}^{N}
\xi_{i}^{\mu}S_{i}(t)
\end{equation}

ii) The storage capacity of the system;

\begin{equation}
\alpha = \frac{P}{N_{Synaps}}
\end{equation}

where $N_{Synaps}$  is the number of remaining connections per neuron and 
$P$ is the number of stored pattern in the diluted network.
In this work we considered a fixed number of connections for each neuron.

\section{Results and Discussions}

In this work we have employed the zero temperature dynamics,
Eq. (\ref{Dynamics}), for the time evolution of the neurons. We have
used four different-size neuron networks and for each one of these
networks different dilution rates are employed. The number of neurons
$(N)$ and the number of connections per neuron $(N_{Synaps})$ that are
used in this simulation study are given in Table 1.

$\;$

In our studies the initial patterns are prepared for each size, $N$,
by using random linear distribution. The set of $P$ initial patterns
are stored into the system. The connections between the neurons are
calculated by using the above given prescription until the required
dilution is reached. The patterns are distorted by using $10\%$ random
noise. In our simulation work we have tested the recovery for each of
the distorted initial patterns and this process is continued until all
of the patterns are exhausted. The recovery rate is defined as the
average of the overlaps between the original pattern and the final
recovered pattern.

$\;$

In order to obtain the critical value of storage capacity, $\alpha_c$,
we have started with very small number of stored patterns to our
simulation process and we have increased the number of initial
patterns for each $N$ and $N_{Synaps}$ until no more patterns can be
recovered. This process of feeding the system with increasing number
of initial patterns is continued while $100\%$ overlap between the
final patterns and any one of the original patterns is obtained.

$\;$

In Fig.1 we have presented the storage capacity
($\alpha=\frac{P}{N_{Synaps}}$) versus the number of recovered
patterns for $N =500$ neuron system. Starting from the fully connected
system ($N_{Synaps}\; =\;499$), the behaviour of the diluted system
with varying number of connections per neuron ($N_{Synaps}\;=\;300,
200, 100, 50$) is presented. For a given number of synapses the number
of stored patterns exhibits a peak. This peak becomes wide
Gaussian-like and the maxima of the peaks are shifted toward the
larger $\alpha$ values as the number of connections is reduced. The
initial increase is linear and the slope of the linear initial
increase decreases as the number of synapses decreases. The maximum
number of stored patterns also decreases but the number of stored
patterns per synapse ($P/N_{Synaps}$) increases with decreasing number
of synapses. There is a peak at a given $\alpha$ value and number of
recovered patterns decreases as $\alpha$ increases for each
$N_{Synaps}$. We have applied the linear regression technique to the
decreasing side of each Gaussian-like curves to determine the
critical storage capacities of the network for different number of
synapses per neuron. These calculated storage capacities are used to
determine the critical storage capacity for the fully connected and
extremely diluted systems. These values are presented in Fig.4.

$\;$

In Fig.2, we have presented the storage capacity $(\alpha)$ versus the
average overlap $(m)$ between initial patterns and the recovered
memory for a network of $500$ neurons. Different curves represent
different $N_{Synaps}$ values. While the number of synapses decreases
($N_{Synaps} = 250, 100, 50, 25, 20, 15, 6$) the storage capacity
increases.  As it can be seen from Fig.2, for $N_{Synaps} = 250$, the
storage capacity is already approximately twice the value of $0.14$.
As the number of synapses is decreased to $N_{Synaps} = 6$ this
storage capacity reaches to an average value of $1.35$.  For a system
with $500$ neurons $6$ synapses per neuron seem to be the limit of the
possible dilution.

$\;$

In Fig.3, we have presented the storage capacity
$(\alpha)$ versus the overlap $(m)$ between initial patterns and the
recovered memory for the network of $N = 400, 500$ and $800$
neurons. Here we have plotted $N_{Synaps} = 160, 50$ and $6$ connections
per neuron for each different-size network. This figure shows the
relation between the dilution rate and finite-size effects. The size
dependence of the storage capacity diminishes as the dilution is
increased. As the system becomes extremely diluted, the different-size
networks yield the same storage capacity. In our case ($N = 400, 500,
800$), different network sizes yield the same critical storage capacity
within the error bars which indicate that all curves follows the same
path from the storage to the total memory loss phase.

$\;$

In order to study the limiting cases, we have presented the relation
between the fraction of diluted synapses and the critical storage
capacity. In two extreme regions we have fitted the data to the same
form,

\begin{equation}
\alpha_c(F_d) = \frac{a_{0}}{1-a_{1}F_d}.
\end{equation}

where $F_d$ is the fraction of diluted synapses, $F_{d} = 1 -
\frac{N_{Synaps}}{N}$ and $\alpha_c(F_d)$ is the critical storage
capacity at given dilution fraction.  In Fig.4, the dilution fraction
versus critical storage capacity is presented for the networks
$N=400,500,\;{\rm and}\;800$. The data and the fits to the data for
the largest lattice, $N=800$ at two extreme limits ($F_{d} \rightarrow
0$ and $F_{d} \rightarrow 1 $) are also presented in this figure.  In
both limits the same functional form for the fit (Eq. 9) has been used.  
In the first limit (fully connected system) fit yields the Hopfield model critical
storage capacity value as expected.  The second region ($F_{d}
\rightarrow 1$) includes data of at least $ 90 \% $ diluted system as
seen in Fig.4. Fit to the data of the diluted synapse ratio $F_{d} >
0.9$ gives the critical storage capacity which is approximately equal
to $\alpha_{c}(1) = 1.6$.

\section{Conclusion}

In this work we have studied the storage capacity of extremely diluted
Hopfield Model. Our dilution algorithm is based on a prescription  which
retains maximum information on the non-diluted connections.  This is
achieved by considering the sites which will be connected to the
originally chosen neurons as the ones with the highest number of
patterns containing like and unlike neurons. This choice ensures
the sign of the $i^{th}$ neuron for the next time step. Decreasing
number of connections increases the storage capacity of the system. In
our study we have calculated the critical storage capacity of the
extremely diluted Hopfield Model as about $1.6$.

$\;$

With the well-defined dilution prescription given above and the rate
of increased  storage capacity, our method of dilution open
possibilities for many applications. Particularly at the $100\%$
recovery region, a large amount of information can be stored and
recovered at a very low cost of memory space. Our studies in this
direction is in progress.

\section*{Acknowledgments}

We gratefully acknowledge  M. Ayd{\i}n for the
illuminating discussions and reading the manuscript. The project is
partially supported by Hacettepe University Research Unit (Project no: 01
01 602 019) and Hewlett-Packard's Philanthropy Programme.

\pagebreak

\pagebreak

\section*{Table captions}

Table 1 The size of neuron systems and the number of connections per neuron.

\pagebreak

\section*{Figure captions}

{\bf Figure 1} {The storage capacity ($\alpha$) versus number of recovered patterns 
for $N_{Synaps} = 500, 300, 200, 100, 50$ and linear regressions.}\\

{\bf Figure 2} {The storage capacity ($\alpha$) versus average overlap ($m$) for 
$N_{Synaps} = 250, 100, 50, 25, 20, 15, 6$.}\\

{\bf Figure 3} {The storage capacity ($\alpha$) versus average overlap ($m$) for
$N = 400, 500, 800$ and $N_{Synaps} = 160, 50, 6$.}\\

{\bf Figure 4} {The fraction of diluted connections ($F_{d}$) versus the 
critical storage capacity ($\alpha_{c}(F_{d})$) for $N = 400, 500, 800$.}\\

\pagebreak

\begin{table}
\begin{center}
\begin{tabular}{c l}\\
\hline
Network Size  ($N$) &Number of Couplings per Neurons ($N_{Synaps}$)\\
\hline
\hline
400&400, 360, 320, 280, 240, 200, 160, 120, 100, 80, 50, 40, \ldots, 6\\
500&500, 450, 400, 350, 300, 250, 200, 160, 150, 100, 50, 25, \ldots, 6\\
800&800, 720, 640, 560, 480, 400, 320, 240, 160, 80, 60, 50, \ldots, 6\\
\hline\\
\end{tabular}
\centerline {Table 1}
\end{center}
\end{table}

\pagebreak

\begin{figure}
\vspace*{13pt}
\centerline{\includegraphics[angle=0,width=10truecm]{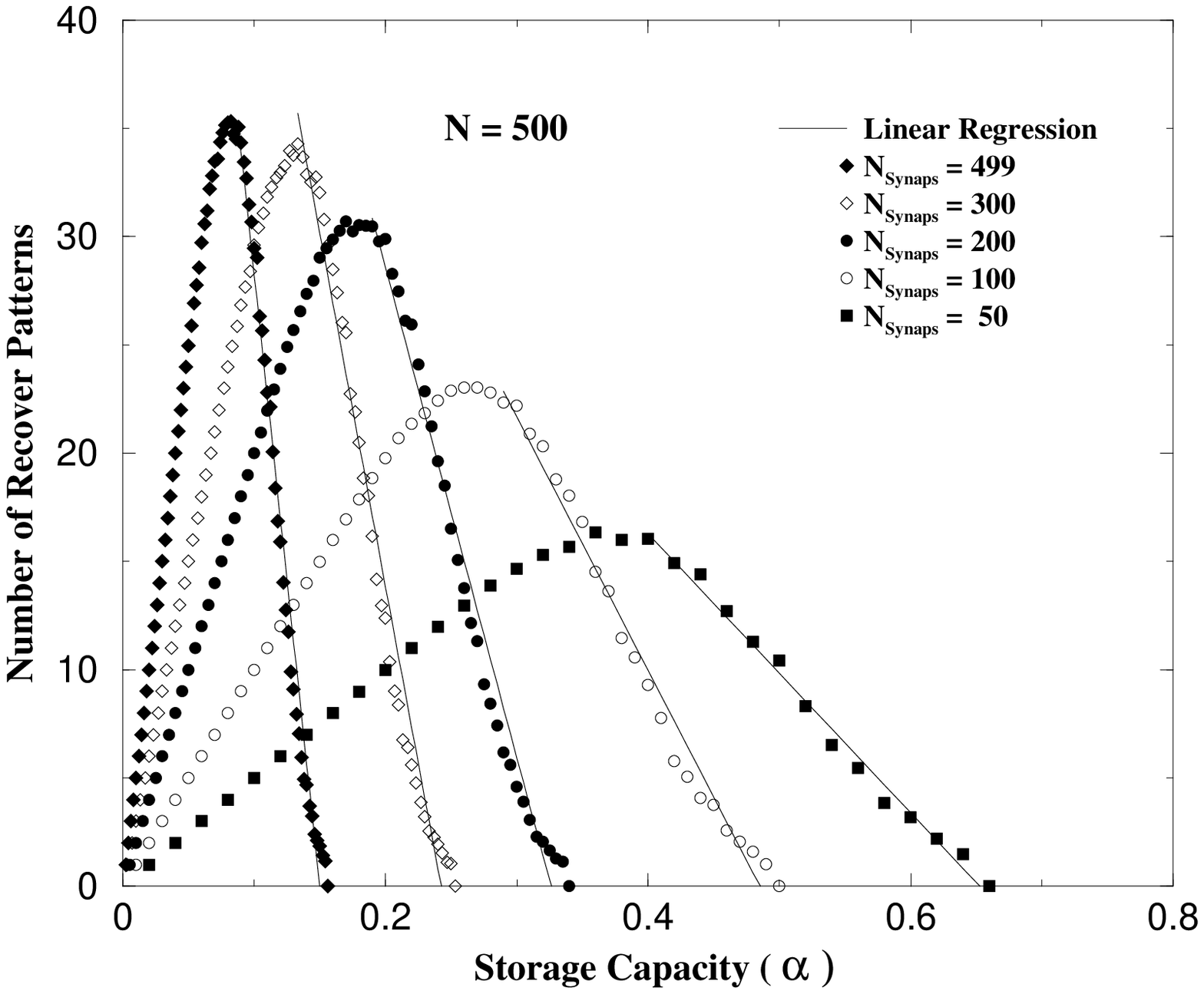}}
\vspace*{13pt}
\centerline {Figure 1}
\end{figure}

\begin{figure}
\vspace*{13pt}
\centerline{\includegraphics[angle=0,width=10truecm]{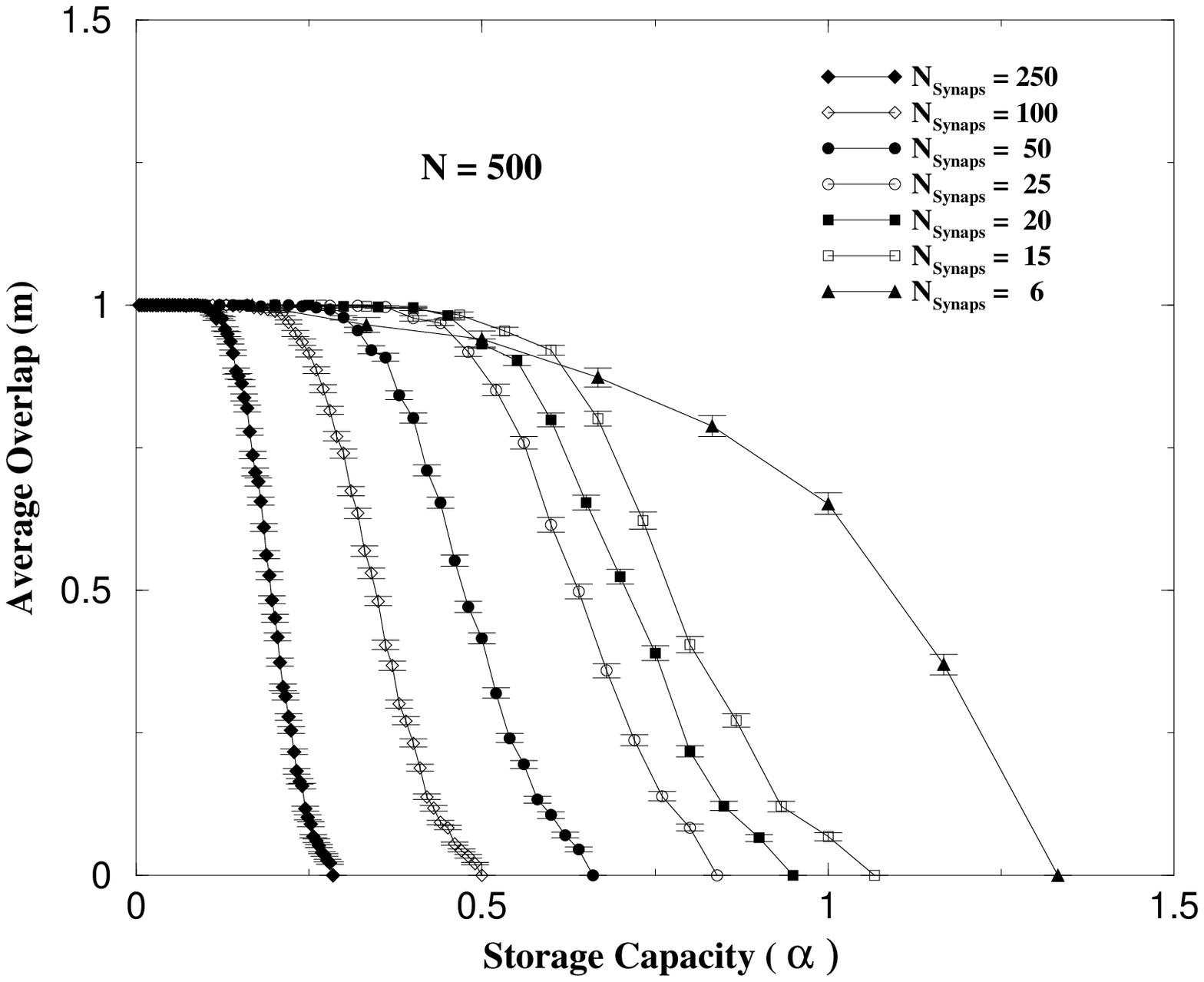}}
\vspace*{13pt}
\centerline {Figure 2}
\end{figure}

\begin{figure}
\vspace*{13pt}
\centerline{\includegraphics[angle=0,width=10truecm]{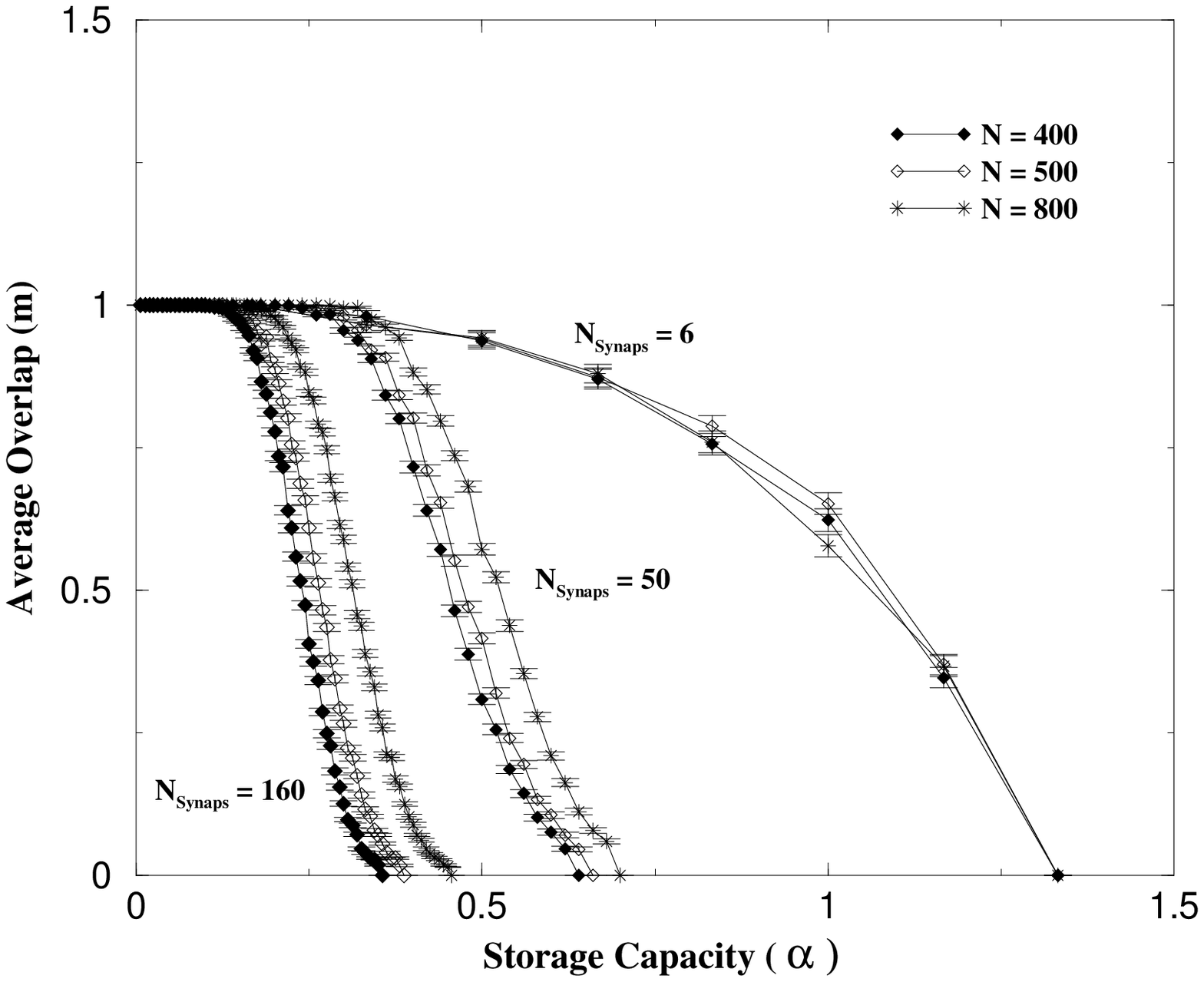}}
\vspace*{13pt}
\centerline {Figure 3}
\end{figure}

\begin{figure}
\vspace*{13pt}
\centerline{\includegraphics[angle=0,width=10truecm]{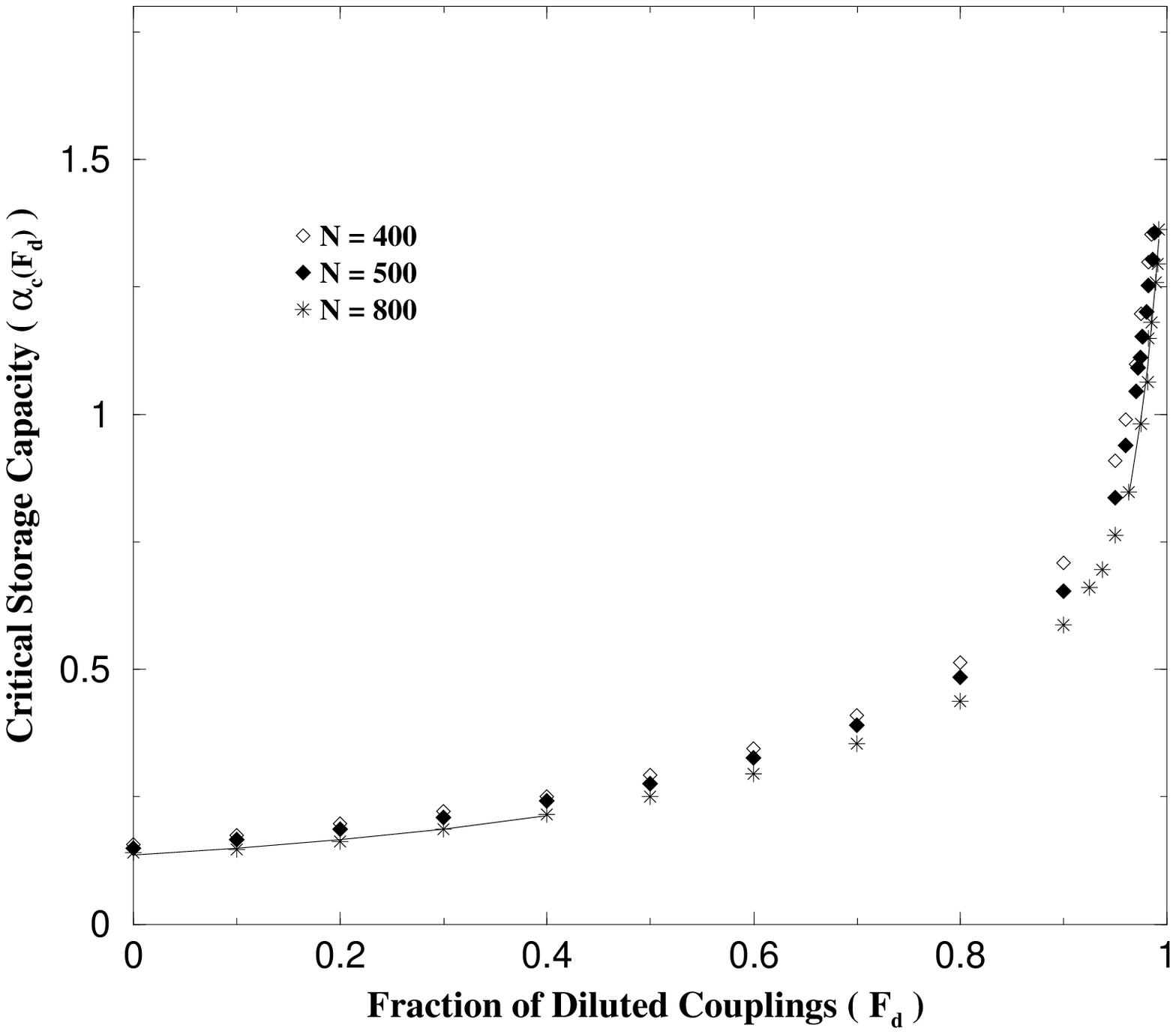}}
\vspace*{13pt}
\centerline {Figure 4}
\end{figure}

\end{document}